# The influence of van der Waals forces on the waveguide deformation and power limit of nanoscale optomechanical systems

*Bi‑cai Zheng, Wei Luo, Fei Xu*and Yan‑qing Lu**

*feixu@nju.edu.cn and yqlu@nju.edu.cn

National Laboratory of Solid State Microstructures and College of Engineering and Applied Sciences, Nanjing University, Nanjing 210093, P. R. China

KEYWORDS: van der Waals force; optomechanical systems; transverse gradient force; plastic/elastic deformation; stiction; optical power limit

**ABSTRACT:** The ultra-short range force, van der Waals force (VWF), will rise rapidly when one nanoscale waveguide is close to another one, and be stronger than the external transverse gradient force (TGF). We theoretically investigate the giant influence of the VWF on the device performance in a typical optomechanical system consisting of a suspended silicon waveguide and a silica substrate including waveguide deformation stiction and failure mechanism. The device shows unique optically-activated plastic/elastic behaviors and stiction due to the VWF. When the input optical power is above the critical power, the waveguide is sticking to the substrate and the deformation is plastic and unrecoverable, even though the total force is less than the yield strength of the waveguide material. This is important and helpful for the design and applications of optomechanical devices.

PACS numbers: 34.50.Dy, 42.50.Wk, 68.60.Bs

Optomechanical system is a burgeoning field that combines nanophotonic and nano-electromechanical devices seamlessly in an integrated system. Various configurations have been proposed and demonstrated in optomechanical systems including high-fineness cavity with radiation-pressure force[1-3] and evanescently coupled optical waveguides with the transverse gradient force (TGF). In the evanescently coupled configuration, a transverse gradient force can be generated from the lateral gradient of a propagating light field in nanoscale waveguides and devices[4, 5]. The force can be used directly for electromechanical actuation and precisely control the positions of optical waveguides and devices[6-14]. This TGF works with the highly-evanescently nanoscale photonic-wire and does not require a reflective surface, thus it is more versatile for future large-scale photonic integrated circuits in planar structures than the radiation pressure force. So far, a number of devices have been studied for realizing the TGF between a substrate and a suspended waveguide or two parallel waveguides. These novel devices may find plenty of applications in optical signal processing, which include nonlinear signal processing devices[8], tunable directional couplers[10], tunable birefringence devices[11], wavelength -filtering devices[12] etc.

In order to make as much use of the force as possible, the width of the waveguide and the gap between the waveguides or substrates should be as small as possible. In general, the size is of tens or hundreds of nanometers. Tens of μ N attractive or repulsive force can be achieved at the input optical power of tens of mW[10, 11]. Smaller gap between waveguides means strong evanescent field and the TGF. However, another type of force, the internal force--van der Waals force (VWF), which is sometieme referred as the Casimir force in the microelectromechanical system[15], arises and can be relatively strong at such small sizes compared to the TGF. The VWF originates from the instantaneous dipole generated by the fluctuation of electron cloud surrounding the nucleus of electrically neutral atoms[16]. It is an ultra-short range force and decays rapidly to zero away from the surface. Different with the TGF, the VWF is generally attractive in nature, and has nothing to do with the optical power. In fact the VWF which is stronger than the TGF at low power can exist between optical waveguides with a nanoscale gap. In previous work of nanoscale optomechanical devices, it is mostly neglected except the calculation of the shot-range Casimir force between two waveguides[17]. It is important to study the competition and cooperation between the VWF and the TGF and how the VWF affects the physical geometry and optical and mechanical characteristics in optomechanical devices. In this paper, we investigate theoretically the contribution from the VWF in the performance of a typical optomechanical device. We calculated the mechanical deformation induced by different forces. The results show that the VWF plays a very important and different role in optomechanical devices. A significant influence by the VWF is that the waveguide will show distinctive optically-activated plastic/elastic behaviors and stiction. When the optical power is relatively low, the waveguide deformation is elastic and can return to its original shape if the optical power decreases to zero. However, there is an elastic limit related to the input optical power due to the existence of the VWF. The waveguide deformation resembles plastic deformation and the waveguide will be permanently bonded to the substrate when the optical power increases beyond the critical value. The phenomenon can occur even though the waveguide material is still in its elastic region. These novel and unique characteristics are important and helpful for the design and operation of optomechanical devices, and would possibly open new routes to many potential applications such as switch and memory.

We consider the typical structure in optomechanical devices as shown in Fig. 1. It is a simple and generic configuration which consists of a silica substrate and a suspended silicon waveguide.

The waveguide is designed to separate from the substrate by a gap of size $g$, in the absence of deformation. The suspended section has a length of $L$ and the waveguide has a cross section of dimensions $a \times d$. From previous work[6], it is known that the force on the free-standing waveguide arises when the guided light evanescently coupled to the dielectric substrate. The magnitude of the force is proportional to the optical power, and the sign is always attractive. On the other hand, strong intermolecular and surface effects give rise to the VWF even without the input power. That force can deform the beam and thus change the effective refractive index.

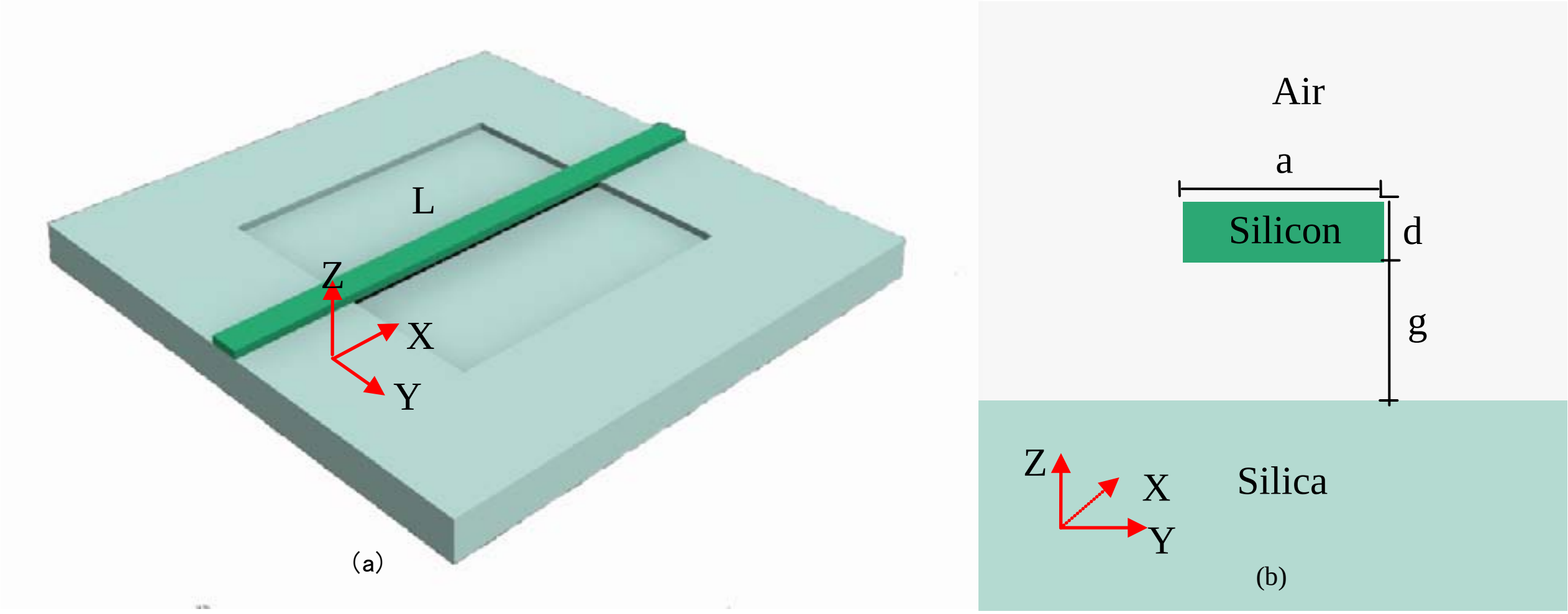

**Figure 1.** (a) 3D illustration of the structure. (b) The cross section view of the structure. The width of the waveguide is $a$ = 500 nm; the height of the waveguide is $d$ = 110 nm; $g$ is the designed gap between the waveguide and the substrate, in the absence of deformation. The drawing is not to scale.

According to the VWF theory, the potential between two small molecules is purely attractive when the distance between molecules is more than several angstroms[16] and its relationship with the distance is $w(r) = -C/r^6$, where $w(r)$, $C$, $r$ represent the potential, the potential constant, the distance between two small molecules, respectively. If we assume that a small molecule interacts with an infinitely large substrate, we can get the corresponding potential $w(D) = -\pi \rho_s/6D^3$, where $D$, $\rho_s$ represent the distance between the molecule and the substrate, the number density of molecules in the substrate[16], respectively. To calculate the potential per unit length between the waveguide and the infinitely large substrate, we have to integral *w(D)* in the waveguide area:

$$w(g)=\int_g^{g+d}\rho_w a w(z)dz=\frac{-C\rho_w\rho_s a\pi}{12}\left(\frac{1}{g^2}-\frac{1}{(g+d)^2}\right)=\frac{-Aa}{12\pi}\left(\frac{1}{g^2}-\frac{1}{(g+d)^2}\right) \quad (1)$$

where $g$ is the gap between the waveguide and the substrate, $\rho_w$ is the number density of molecules in the waveguide, $z$ is the direction perpendicular to the substrate, $A$ is the Hamaker constant between the materials: $A = \pi^2 C\rho_s\rho_w$[16]. We have to mention here that the problem of additivity is avoided in the Lifshitz theory and Eq. (1) is valid even within the framework of continuum theories[16]. The only thing that changes is the way we calculate the Hamaker constant when we apply Eq. (1) in the Lifshitz theory. According to a simplified model, the Hamaker constant between two different materials is $A = (A_1A_2)^{1/2}$ , where $A_1$ and $A_2$ are the Hamaker constant of material 1 and material 2 respectively[18]. For pure silica, the Hamaker constant is $6.5\times10^{-20}$J and for pure silicon, the Hamaker constant is $1.1\times10^{-18}$J[19]. Considering the retardation effect of the VWF at the relatively large separations (>5nm), the full Lifshitz equation should be applied to calculate the Hamaker constant[16]. However, in the gap range in our model (<100nm), we can use the following simplified equation to calculate the Hamaker constant with the retardation effect: $A_{ret} = A_{non\text{-}ret}/(1+pg/100nm)$ , where $p$ is estimated to be around 10 in our model[16].

Then, according to virtual work principle, we can get:

$$F_{VWF}=-\frac{\partial w(g)}{\partial g}=\frac{-A_{ret}a}{6\pi}\left(\frac{1}{g^3}-\frac{1}{(g+d)^3}\right) \quad (2)$$

where $F_{VWF}$ represents the VWF per unit length. Using Eq. (2), we can plot the relationship between $F_{VWF}$ and $g$, which can be seen in Fig. 2a.

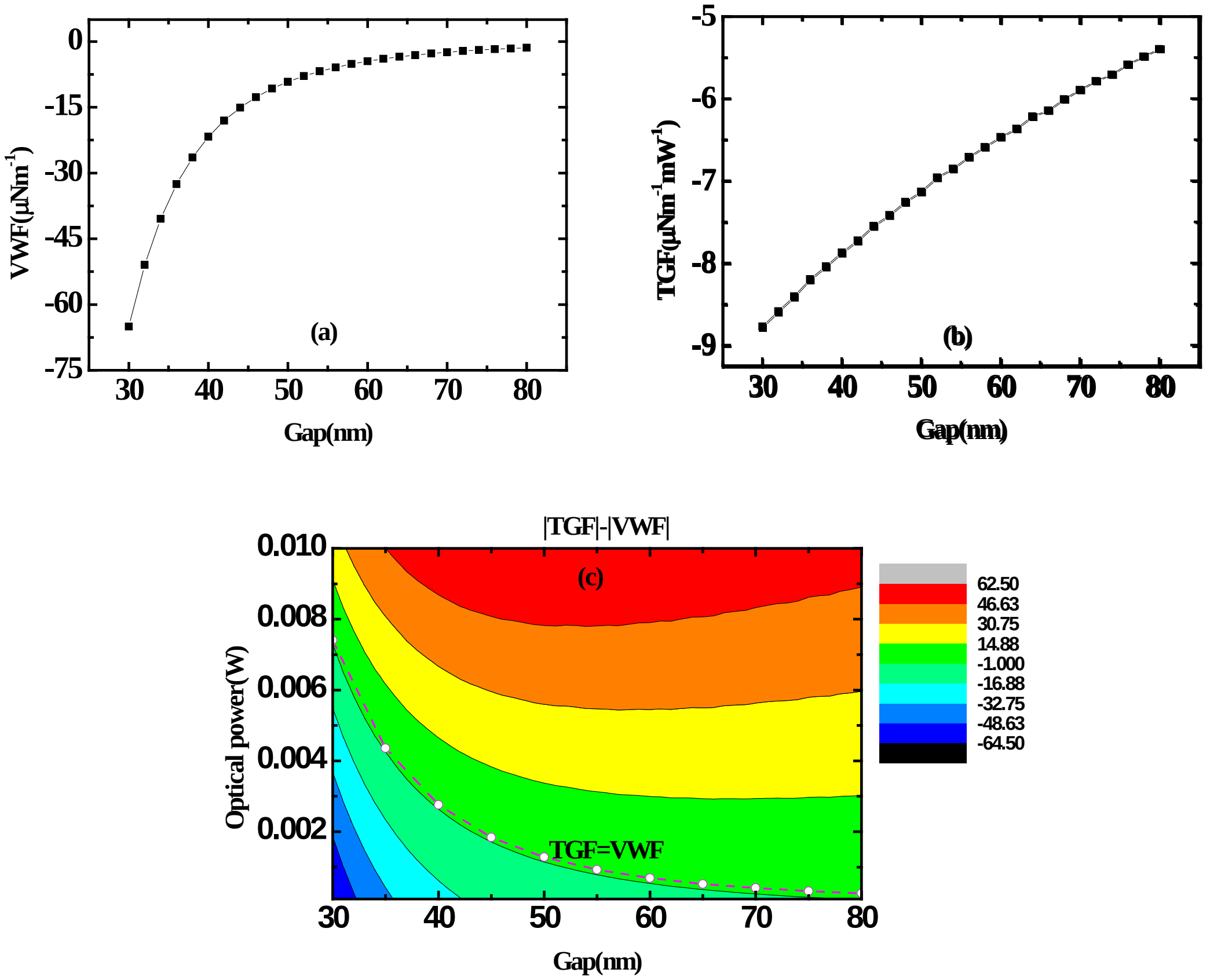

**Figure 2.** (a) The relationship between the VWF per unit length and the gap. (b) The relationship between the TGF per unit length per unit power and the gap. (c) The difference between the VWF and the TGF. The unit is μ $Nm^{-1}$. The dashed line is the power-gap curve when TGF = VWF.

For infinitely long waveguides without loss, the TGF can be calculated as follow:[4]

$$f_{TGF} = -\frac{1}{c}\frac{n_g}{\omega}\frac{\partial \omega}{\partial g}\bigg|_k \tag{3}$$

where $f_{TGF}$ represents the TGF per unit length per unit power, $c$ is the speed of light in vacuum, $n_g$ is the group index, $\omega$ is the frequency and $k$ is the wave vector. By substituting $\omega = ck/n_{eff}$ into Eq. (3), the TGF can be expressed as a function of the effective index $n_{eff}$:

$$f_{TGF} = \frac{1}{c}\frac{n_g}{n_{eff}}\frac{\partial n_{eff}}{\partial g}\bigg|_k \tag{4}$$

Using Eq. (4), we can plot the relationship between $f_{TGF}$ and $g$ when the wavelength is fixed at 1550 nm, which can be seen in Fig. 2b.

From the figures above, the VWF increases more rapidly than the TGF when the gap declines to zero. It can roughly be estimated that when the optical power is around 0.5 mW and the gap is around 80 nm, the VWF and the TGF have the same order of magnitude. To further investigate the relationship between these forces, a new function can be defined as:

$$F(P,g) = P\left|f_{TGF}\right| - \left|F_{VWF}\right| \tag{5}$$

where $P$ is the optical power, $|f_{TGF}|$ and $|F_{VWF}|$ are the absolute values of the TGF and the VWF. The calculated results are plotted in Fig. 2c. The figure shows that when optical power is low and the gap is small (the light green region to the deep blue region), the VWF is comparable with or even much larger than the TGF, which indicates that the VWF can't be neglected and even dominated under this circumstance. The balanced conditions are shown as the dashed line when the TGF equals to the VWF.

The suspended silicon waveguide on the silica substrate can be modeled as a double-clamped beam[20]. When the maximum deformation (*MD*) of the waveguide is much smaller than the length of the waveguide, the deformation along the waveguide is decided by the following deformation equation:[20]

$$EI\frac{d^4u(x)}{dx^4} = -q(x) \tag{6}$$

where $E$ = 131 Gpa is the Young's modulus of silicon, $I = ad^3/12$ is the moment of inertia of the waveguide, $x$ is the direction along the waveguide, $u(x)$ is the deformation distribution along the waveguide and $q(x)$ is the force per unit length along the waveguide.

If we want to get the accurate solution, a numerical Iterative Feedback Tuning method has to be employed and it will be discussed in the following paragraphs. Here we try to obtain the approximate solution by a simplified averaged-force method. We firstly consider the simplest case of a uniform load, and then an analytical solution can be given:

$$u(x) = -q_0 x^2 (x-L)^2 / 24EI = -MD(2/L)^4 x^2 (x-L)^2 \qquad (7)$$

The *MD* occurs at the center of the waveguide. The relationship between uniform load $q_0$ and the *MD* is shown in Fig. 3 (the black line). Here the designed gap between the waveguide and the substrate is fixed at 80 nm and the length of the waveguide is 30 μ m. We can calculate the *MD* if the load is beyond the yield strength of the waveguide material[21] and the result is much larger than 80 nm.

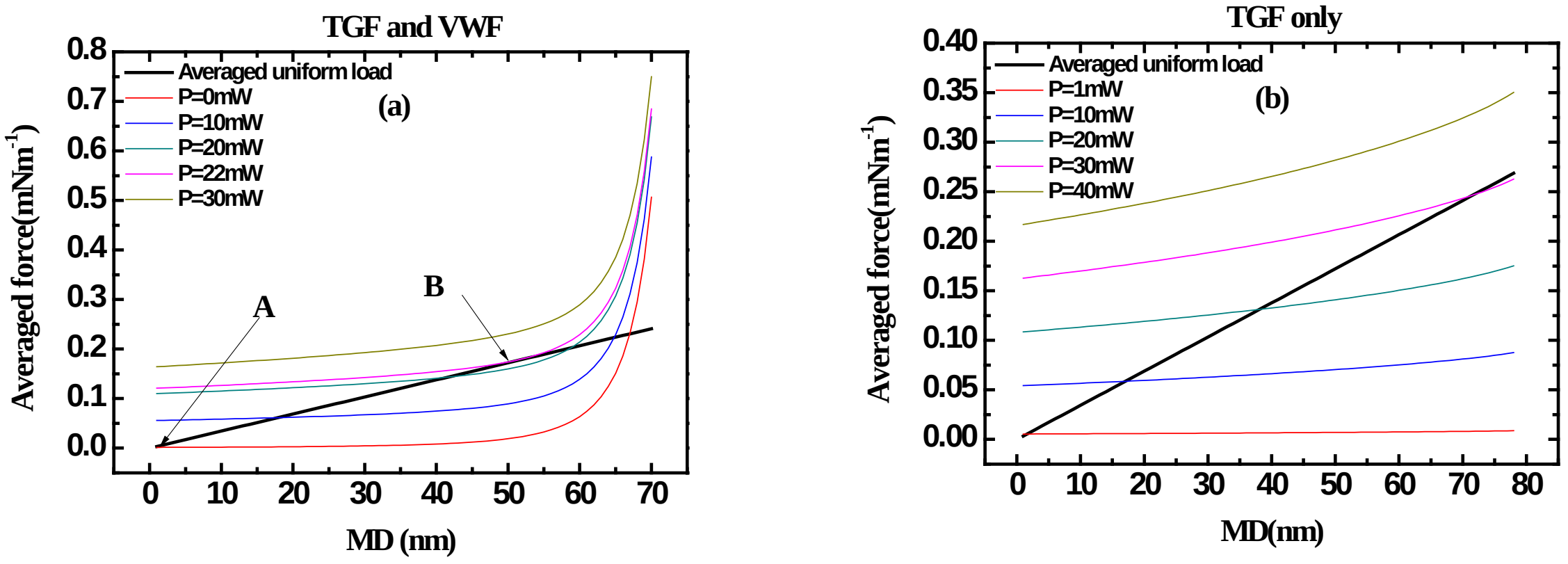

**Figure 3.** The force-deformation response at the simplest case of an assumed uniform load (the black lines) and the averaged force for different optical input power at a specific deformed shape $u(x) = -MD(2/L)^4x^2(x-L)^2$ (the colored lines) when considering (a) TGF+VWG and (b) TGF only. In (a), the left intersection is the approximate stable solution at specific optical power. The interaction A represents the original shape and *MD* only by the VWF (the optical power is zero). B is the critical point above which the black line can't cross the colored line and there is no stable solution.

On the other hand, the averaged force including the VWF and the TGF at a specific deformed shape $u(x) = -MD(2/L)^4x^2(x-L)^2$ can be calculated using Eq. (2) and Eq. (4). The averaged forces (VWF+TGF) of various optical power are shown in Fig. 3a (the colored lines).

There are two intersections between the black and one colored line at low optical power. The right intersection with larger *MD*, corresponds to an unstable equilibrium state. At the left intersection with smaller *MD*, a stable equilibrium state with such a *MD* can be realized by the VWF and the TGF at corresponding optical power. This is an approximate solution. The interaction A in Fig. 3a represents the original shape and *MD* only by the VWF (the optical power is zero). The static deformation increases with optical power and the deformation is elastic because the waveguide can return to its original state at A after the TGF is removed.

However, the two intersections approach when the optical power increases and coincide each other (B in Fig. 3a) at a critical power. The black line can’t cross the colored line above B. When the optical power is above the critical power ($P_c$) at B, the force is too large to stabilize the deformation, and then the beam is pulled down to the substrate and the middle part of the beam is bonded on the surface. The final *MD* is as large as the designed gap (80 nm). More parts of the beam will be collapsed into the substrate surface if the optical power further increases. However the deformation is plastic and irrecoverable, and the waveguide is permanently bonded on the substrate by the ultra-strong VWF, even after the TGF is removed.

If we only consider the TGF as shown in Fig. 3b, there is also a critical power $P_c$. When P > $P_c$, there is no stable solution and the waveguide will be pull down to the substrate. However, due to no VWF, it is always elastic and can return back after the optical power decreases to zero.

The optically-activated plastic/elastic behavior is very interesting, but it also limits the power lever in the waveguide. The above method can be implemented to roughly obtain the approximation of the critical power. In order to get the exact value, the Iterative Feedback Tuning method has to be employed to get a convergence result. Using this method, we accurately calculate the *MD* at the center of the waveguide under different input optical power. Figure 4a shows the evolution of deformation shape versus the optical power by the total force (TGF+VWF). When $P = 0$, the waveguide is deflected little only by the VWF, and then deflected more with increased power. As $P$ is at the critical power of 18.3 mW, it achieves a critical state, and it still can be recoverable after the stress is removed and returns to their original state at $P = 0$.

However the waveguide will irreversibly be pulled to the substrate quickly when $P > P_c$. More parts of the beam will be collapsed into the substrate surface if the optical power further increases, and the waveguide can't return to its original shape even though the optical power is reduced to zero. The red area is the plastic deformation range. Here we have to mention that the waveguide material is still in its elastic region.

If we only consider the TGF, there is also a critical power at 25.5 mW, but the waveguide can return to the original state at $P = 0$.

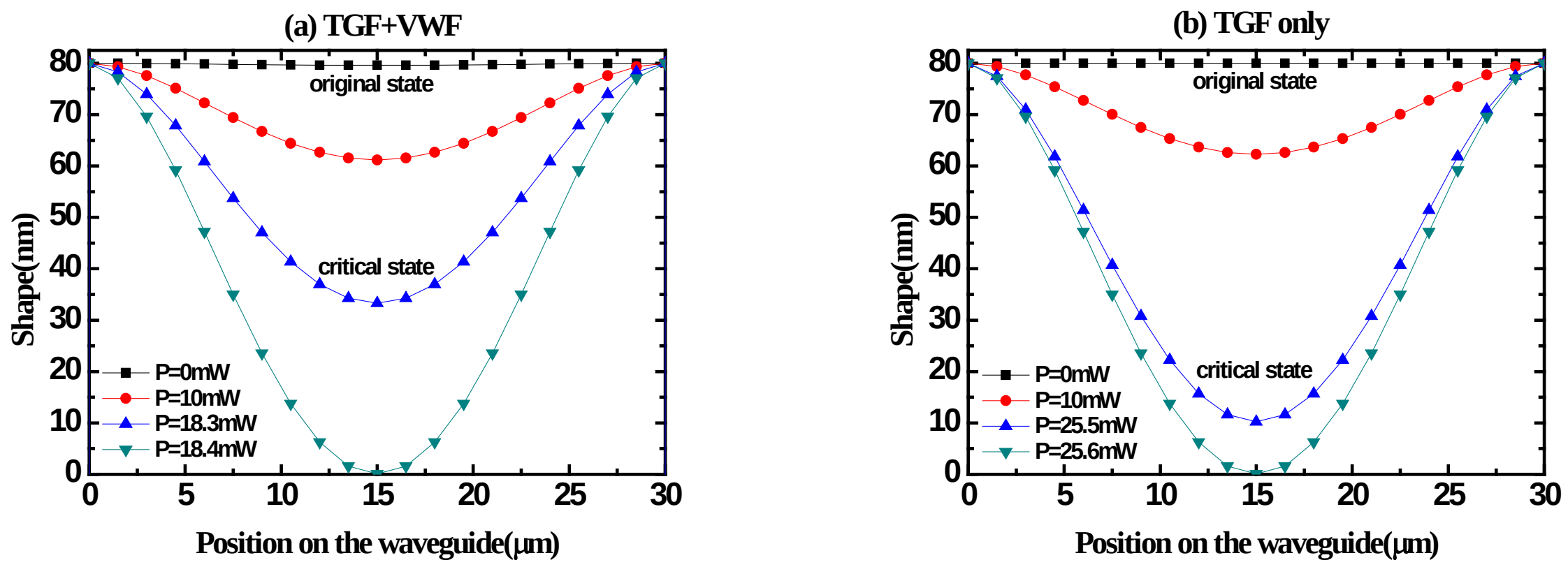

**Figure 4.** The evolution of the deformation shape versus the optical power by (a) and the total force (TGF +VWF) and (b) TGF only. The white and red areas are the elastic and plastic deformation range, respectively.

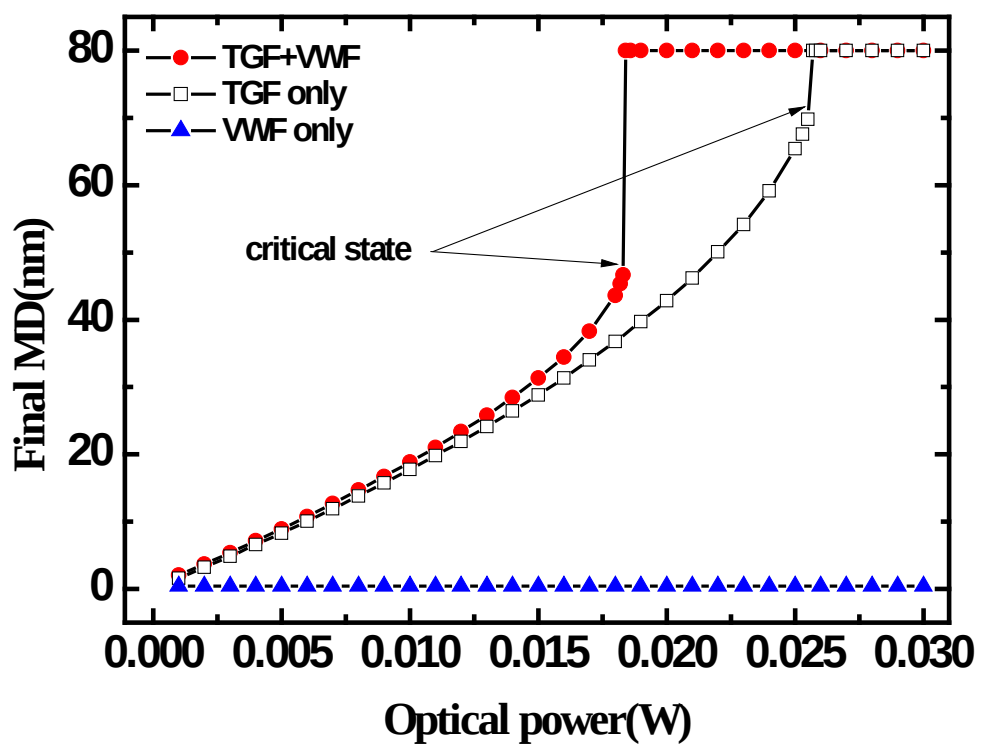

**Figure 5.** The relationship between the final *MD* and the optical power for TGF, VWF, and TGF+VWF, respectively.

Figure 5 shows the curve of the final *MD* versus the optical power, for the TGF, the VWF, and TGF+VWF, respectively. The final *MD* for TFG+VWG is almost linear to the optical power if $P<<P_c$. Another thing has to be mentioned is that the difference between $MD_{TGF+VWF}$ and $MD_{TGF}$ $+MD_{VWF}$ is always positive, and increases when the optical power increases.

We also calculate the critical power at different waveguide lengths for TGF+VWF. As shown in Fig. 6, $P_c$ is higher with shorter waveguide and it has a nearly linear growth when the gap is above a critical value.

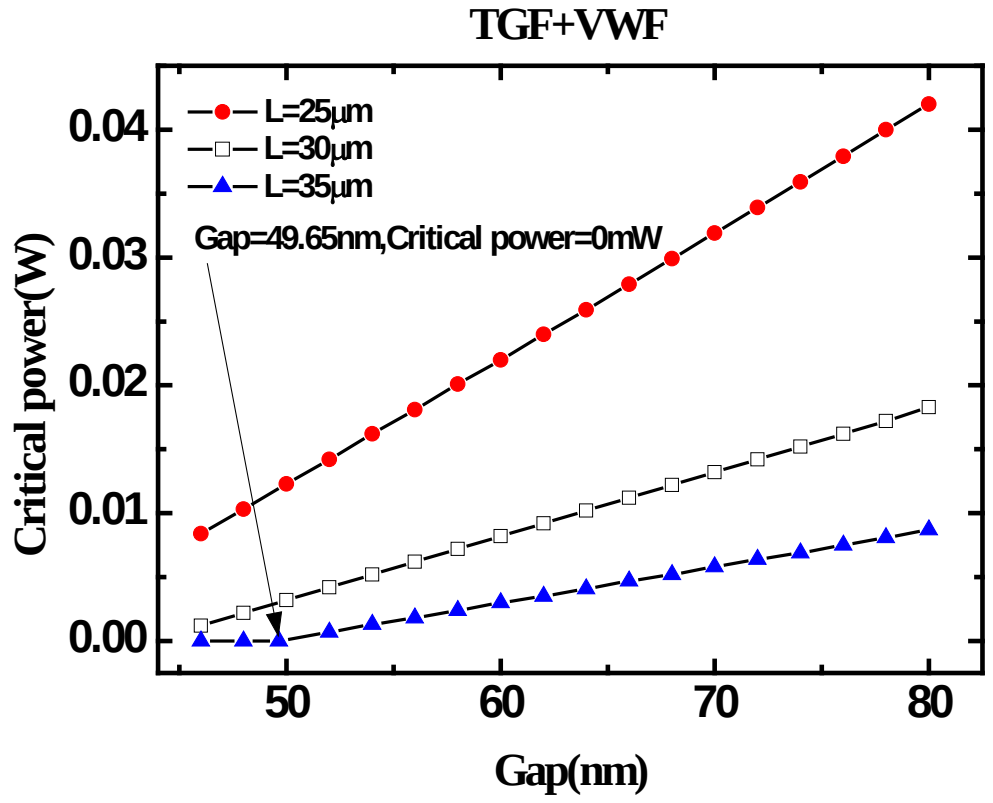

**Figure 6.** The critical power at different waveguide lengths for TGF+VWF. The horizontal part of the blue dashed line means that when the gap is smaller than a critical value (which is 49.65nm for *L*= 35μ m), the VWF will straightly pull the waveguide to the substrate without the existence of the TGF. It can be estimated that there will be the same phenomenon for $L$= 25μ m and $L$= 30μ m.

In this paper, we theoretically investigate the VWF's influences on the optomechanical system consisting of a suspended silicon waveguide and a silica substrate. The calculation results show that the VWF should be taken into account in practical design and applications, and it has a great influence on the device performance by deforming the waveguides and thus tuning their effective refractive indices. When the input optical power is relatively low, the VWF is comparable with the TGF in magnitude; when the input optical power is relatively high, and the interaction between the VWF and the TGF may make the VWF even larger than the TGF after the deformation.

Due to the unique property of the VWF, the waveguide deformation has the plastic/elastic behaviors depending on the input optical power. When the input optical power is above the critical power, the waveguide deformation is plastic and unrecoverable, even though the total force is less than the yield strength of the waveguide material. The interesting phenomena possibly pave a way for the new applications of the optomechanical devices. Finally the results obtained in this work are also important and helpful for the design and operation of optomechanical devices.

## Funding Sources

This work is supported by National 973 program under contract No. 2012CB921803 and 2011CBA00205, NSFC program No. 11074117 and National Science Fund for Excellent Young Scientists Fund (61322503) and National Science Fund for Distinguished Young Scholars (61225026). The authors also acknowledge the support from PAPD and the Fundamental Research Funds for the Central Universities.